# Analysis, Detection, and Classification of Android Malware using System Calls

Shubham Shakya[1] and Mayank Dave[2]
Department of Computer Engineering
National Institute of Technology
Kurukshetra 136119 Haryana India
{[1]shubby98@gmail.com, [2]mdave@nitkkr.ac.in}

**Abstract**

With the increasing popularity of Android in the last decade, Android is popular among users as well as attackers. The vast number of android users grabs the attention of attackers on android. Due to the continuous evolution of the variety and attacking techniques of android malware, our detection methods should need an update too. Most of the researcher's works are based on static features, and very few focus on dynamic features. In this paper, we are filling the literature gap by detecting android malware using System calls. We are running the malicious app in a monitored and controlled environment using an emulator to detect malware. Malicious behavior is activated with some simulated events during its runtime to activate its hostile behavior. Logs collected during the app's runtime are analyzed and fed to different machine learning models for Detection and Family classification of Malware. The result indicates that K-Nearest Neighbor and the Decision Tree gave the highest accuracy in malware detection and Family Classification respectively.

## 1. Introduction

Google Android owns almost 75% market share of smartphones with 2.8 billion active users [1]. Its open-source nature and simplicity to customizability at different levels attracted users and gained manufacturers' attention to producing low-cost smart devices. Android is also popular among the developers' community because of its SDK, which can help developers develop Android applications with minimal effort to hit such a large target. Due to its larger community of users and its popularity Activity of attackers targeting android via malware apps has increased significantly in recent years. According to data from AV-Test, the number of total malware has increased from 182 million to 1342 million in the last ten years [2]. The purpose of a malware app is to do something unwanted, which compromises your privacy and sensitive data. So it is vital to develop a method that can efficiently detect malware apps out of legitimate.

Although a lot of work is done under Android Malware detection, as our detection method evolves over the years, the attackers adopt new practices to invade security. Different approaches have been considered to detect malware. We can broadly classify them into three main categories Static, Dynamic, and combination of these two called hybrid approaches.

The static approach is based on the reverse engineering of apk acronym for the android application package file. It contains AndroidManifest.xml and classes.dex file can extract information like different requested permissions during installation and runtime and API calls. Static analysis requires fewer efforts and is fast. But static approaches are precise only in limited space because they originate from code abstraction.

---
[1] Corresponding Author

The dynamic approach, also known as behavior analysis, is based on the runtime behavior of an application. This analysis can be found on different parameters like Network logs, a list of open files, API calls, and system calls. These data collected during the application's runtime are collected later as logs and followed for malware behavior. This identification of patterns can detect the new behavior. Dynamic analysis is practical but requires more time, computation, and space.

## 2. Related Work

The work done by researchers in this area is continually evolving with time and the techniques adopt new changes according to any new malware. In the literature, various authors have proposed techniques for detecting malware using static analysis and dynamic analysis using different runtime behavior. Zhou et al. [3] discussed the systematic characterization of android malware based on their installation method, activation mechanics, nature of carried malicious payload, and how malware is evolving with time. Sato et al. [4] discussed an approach to detect malware applications based on analysis of manifest files of android apk. This approach extracts some specific information from the manifest file and matches information with a list of keywords. Feizollah et al. [5] discussed the value of application intents as a distinguishing feature for detecting malicious applications. A different approch based on the analysis of android application using MD5 hash [6] is dissused by Dunhum et al. This hash is use later to analyize with similar another hash of apps.

The main drawback with static methods is that they can easily contend with the addition of some irrelevant features and hiding information which is the core of detection strategies. Some recent works with more advanced strategies like using Behaviour analysis and machine learning models are taken into consideration to effectively detect the malware. MADAM [7] discussed an approach to detect malware with multi-level (Package, User, Application, Kernel) behavior-based features on runtime. This approach can effectively detect malware with more than 96 percent accuracy. DyanLog [8] is an automated dynamic analysis-based framework for Identifying android applications that stimulate the user behavior by sending random events from MonkeyRunner [9] and extracting API calls as logs, which is used as a dynamic feature to detect the malware. The drawback with this approach is that we are not sure whether we can trigger that malicious behavior of the application with those events or not. IntelliDroid [10] is a Target Input Generator for the Dynamic Analysis that generates a small set of inputs from static analysis of apk that will trigger almost every part of the code. Apart from all the above approches a slightly different approch using MalDy Framework [11] based on analysis of behavioural report of malware. MalDy Framework is built on the top of Natural Language Processing modeling and supervised machine learning methods.

But to bypass, some advanced methods that malware writers are using to hide malicious behavior are not enough. Techniques like Obfuscation of code and dynamic compatibility to download and execute the payload at runtime [3]. These advanced methods can behave differently at the API level in other applications, making it difficult to make out patterns at the API level. Hammad el at. [12] discussed an effect of obfuscation technique on android apps, study shows this technique can be used to confuse malware detectors. Malik et al. [13] discussed an analysis of system calls of android malware families and observed some system calls are more frequent in malicious. Canfora et al. [14] discussed an approach to using monkey tools to interact with applications at runtime to extract system calls of applications. These low-level traces of system calls are fed to the Machine learning model to detect the Malware. Casolare el at. [15] using a similar approach to extracting system call with an automated interaction tool and changing extracted system calls to image representation and using a convolution Neural Nets to detect the malware.

The main focus of this research is to fill the literature gap in dynamic analysis with system calls that can be used without concerning the rapidly updating android technology, emphasis the facts that come out from the analysis of data, and evaluate the data by detecting and classifying the Android malware.

## 3. Background of Android and Android Malware

### 3.1 Android Architecture and Applications

Android is an open-source platform. It is created on the top of the Linux Kernel. It consists of Linux based operating system, a Hardware Abstraction Layer known as HAL, Native Libraries, Android runtime, an application framework, and the topmost layer as System apps [16]. Android Applications are archive files that contain all the of an app. It is also known as the android package, where the extension of the application derived apk. An Android package file includes five main components: Activities, Services, Content Provider, Broadcast Receiver, and Intents [17]. An activity is a screen where users interact with that application and exchange responses. An app contains multiple activities to commute with different parts of an application. Services are the background processes that perform all the computation and execute the operation. A Content Provider is used to perform any function related to data, like exchanging data between different content providers or accessing databases, files, and other resources. A Broadcast receiver is a response to an event by announcing that event system-wide such as a new message. Intents are the interface used to commute between different apps or components of a single application.

An android application also contains the Manifest file, the most critical file. It includes all information required to run an app, like metadata, App components, and Permissions [18]. Metadata contains some standard info of the app like the package name, app id, SDK version, package version, etc. Permissions are an essential piece of information in the manifest file. All the resources which an app will use should be declared here. These permissions can be anything from API of System resources to Access the hardware. These access to this permission while installing or while using that resource later.

### 3.2 System Calls

System calls allow applications to communicate with the operating system's kernel. A user cannot directly communicate with the kernel level. System calls provide the functionality to use the kernel resources for the user via an API; these special APIs are known as System calls. [19] When an application requests a resource via a system call, controls are transferred to privileged kernel mode from unprivileged user mode. This separation of software from hardware via the kernel layer protects the unprivileged application from exploiting the system resources.

System calls are not directly invoked; they are called via a wrapper function. Wrapper functions are sub-level of interface—this wrapper function returns an error by a negative value instead of returning values from hardware. The wrapper function also checks for suitable out of two related system calls, for example, truncate() and truncate64(). In kernel 3.7, there exist 393 system calls [20], out of which a list of some important calls is given in Table 2.1 mentioning the functionality.

**Table 2.1:** List of a few system calls with description [19]

| System call | Description |
| --- | --- |
| Accept | accept a connection on the socket |
| access, faccessat, faccessat2 | check user's permissions for a file |
| bind | a name to a socket |
| chmod, fchmod, fchmodat | change permissions of a file |
| close | a file descriptor |
| connect | initiate a connection on a socket |
| fstat | get file status |
| getuid, geteuid | get group identity |
| kill | send a signal to a process |
| open, openat, creat | open and possibly create a file |
| read | read from a file descriptor |
| recv, recvfrom, recvmsg | receive a message from a socket |
| send, sendto, sendmsg | send a message on a socket |
| epoll_wait, epoll_pwait, epoll_pwait2 | wait for an I/O event on an epoll file descriptor |
| Ioctl | control device |

**3.3 Android Malware**

Android is the most extensive base of users and has a big pool of data. This data can be breached via different methods via using Android malware. This malware is systematically divided into some categories based on the characterization based on its installation, activation, and way of carrying a malicious payload [3].

**3.3.1 Malware Installation and Activation**

Based on the installation of android malware, malware is generalized into three categories [3]. Repackaging, update attack, and drive-by download.

1. **Repackaging:** In the Repackaging technique, attackers use popular apps, disassemble them, and repackage it with a malicious payload. This new repackaged app is listed on the android market. GoldDream, DroidDream, AnserverBot, and DroidKungFu [6] are some families of malware in this category.
2. **Update Attack:** In this technique, the package is disassembled and enclosed with a component that will download a malicious update during runtime. BaseBridge, DroidKungFuUpdate, AnserverBot, and Plankton are families of this category.
3. **Drive-by Download:** this method uses advertisement in the app that tempts users to download it. Examples of this malware category are GGTracker, Jifake, ZitMo, etc.
4. **Other:** Apart from the above category, many other malware families have individual identities. They do not fall under the same group of categories. For example, the Spyware family keeps track of user info. Some families like Asroot are known for exploiting root, etc.

Android malware can depend upon the event to show its malicious behavior; till that point, malware behaves like other regular apps. According to the observation by yajin [3]. These system events are routine procedures for an android like BOOT_COMPLETED, SMS_RECEVIED, BATTERY_LOW, etc.

### 3.3.2 Malicious Payloads

In this section of malware characterization. Malware is divided by the functionality of the malicious payload [3]. It is divided into four categories: Remote Control, Personal Data stealing, Privilege escalation, and Financial charges.

1. **Remote Control:** Malware converts the infected device into bots for remote usage in this method. The system receives commands from other devices. The malware is programmed to establish a connection with the remote server, and the malware author encrypts the address. For example, Pjapps, DroidKungFu3, Geinimi, etc.
2. **Personal Data Stealing:** In this category of malware, malware continuously searches and gather valuable info of user, including Phone numbers, messages, User sensitive info, etc. All this info collected is uploaded on the remote server. For example, Zitmo, Spitmo, etc.
3. **Privilege Escalation:** As we know, Android is multi-layer architecture on the Linux kernel with many open-source libraries. Some part of this architecture is too vulnerable. This category of malware exploits this vulnerable area of android to serve the execution of malware. For example, RATC(RageAgainstTheCage), Zimperlich, GingerBreak, Asroot, etc.
4. **Financial Charges:** This type of malware subscribes to the premium paid services on the user's behalf without the user's awareness—for example, GGTracker, Zone, RogueSPPush, etc.

## 3.4 Obfuscation Strategies

Obfuscation is a technique to change code into a form that is difficult to read or reverse engineer for humans as well as machines. This obfuscation technique is used by malware authors as well as by benign app developers to protect their property from cloning by making the application difficult to reverse engineer. On the other hand, these techniques are used by malware authors to hide malicious activity. There are many different techniques present for obfuscation that the malware author currently uses. These techniques are systematically divided into 2 categories [] trivial and non-trivial.

**Trivial:** In the trivial method, code is obfuscated without change in bytecode. Some of the obfuscation techniques are
1. **Repackaging (REPACK)**: It involves re-signing of apk with a different signature by unzipping it
2. **Disassembling and Reassembling (DR)**: This technique involves reverse-engineering the application file and reassembling it. While disassembling the application the order of items in the dex file is changed. This obfuscation method can confound those malware detection techniques that use dex files to match the signature of known malicious apps.
3. **Android Manifest transformation (AMN):** this technique involves the change in Manifest file like different permissions listed in files to make a malicious app more likely to be a benign app. This technique can easily bypass the check of the non-malicious apps if the manifest file is the only parameter to judge an application.
4. **Alignment (ALIGN):** This technique changes the cryptographic hash of the application file because it realigns all the uncompressed data in an apk file.

**Non-Trivial:** Non-trivial methods involve a change in the bytecode of the application. A list of some Non-trivial methods is
1. **Junk Code Insertion (JUNK):** Junk involves the addition of code like null operations and comments etc. that doesn't affect the execution of the application. But it is part of the application to confuse the reader.
2. **String Encryption (ENC):** ENC involves the encryption of the string and adds a function that will decrypt the encrypted string during runtime.
3. **Control-Flow Manipulation (CF):** CF involves manipulating the flow of code by adding unnecessary conditions and iterative constructs. As result, it transforms the current flow of the call graph of the app
4. **Member Reordering (MB):** MB involves the change in the order of object variable and method in the dex file.
5. **Identifier and Class Renaming:** This technique involves the changing name randomly of classes, variables & methods of classes, and instances. This transformation changes the method table in bytecode.
6. **Reflection (REF):** REF involves transforming direct method calling into reflective calling using the Java reflection feature. This technique can easily evade static analysis that relies on direct method invocation.

All methods to detect malware using static analysis cannot resist these obfuscation techniques as shown in work by (Hammad el at., 2018).

## 3.5 Machine Learning Classifiers Used

In our case, malware detection can be treated as a classification problem where the prediction of a category from malware or benign. It is the case of binary classification, where we have two categories. On the other hand, Family prediction of malware is also a classification problem but with more than two categories that fall into multiclass classification. Classifiers are the algorithms that classify input data. There is a long list of these classification algorithms however, we describe below the most common and very powerful classifiers as used in this research.

### 3.5.1 Logistic Regression

It is an algorithm used to output the likelihood of the categorical dependent variable. It tries to search for a link between a property and particular results. It is very lightweight and efficient for binary classification problems. It can also be used for multiclass classification using one vs. rest comparison.

### 3.5.2 K-Nearest Neighbors (KNN)
This algorithm is very explanatory with its name; it looks for the k nearest point and predicts the point category on the majority vote. KNN is a lazy algorithm because it uses all data for training during classification. K is an experimental parameter whose different values can have different results.

### 3.5.3 Decision Tree (DT)
A decision tree is a tree-like structure where each branch represents the rule, each node represents the feature, and leaf nodes represent the final results. The tree makes decisions at every node of the tree and moves to a conclusion. A decision tree is a very effective technique with a large feature sample.

### 3.5.4 Support Vector Machine (SVM)
Support Vector Machine is an algorithm used to classify the data point in N-dimensional space with a hyperplane. The dimension of the hyperplane depends upon the dimension of space. The closest data point or support vector affects the position of the hyperplane, which is why it is named Support Vector Machine.

### 3.5.5 Multi-Layer Perceptron(MLP)
MLP is a feed-forward neural network that can approximate any continuous function. Every layer contains many nodes named perceptrons. It is a graph-like structure that makes decisions at each stage.

## 3.6 Evaluation Metrics
The metrics used to evaluate the performance of the machine learning model are as under:

- **True Positive:** Those malware samples which are truly classified as malware by the model are known as True Positive(TP).
- **True Negative:** Those Benign samples which are truly classified as Benign by the model is known as True Negative(TN).
- **False Positive:** Those Benign samples which are classified as malware by the model is known as False Positive(FP).
- **False Negative:** Those malware samples which are classified as Benign by the model are known as False Negative(FN).
- **Precision:** The ratio of correct malware prediction(TP) to the total malware prediction (TP + FP). is known as Precision.
$$Precision = \frac{TP}{TP + FP}$$
- **Recall:** The ratio of correct malware prediction to total malware sample is known as recall. This metric tells us how many malware samples are identified.

$$Recall = \frac{TP}{TP + FN}$$

- **Accuracy:** The ratio of correct prediction to total samples is known as accuracy.

$$Accuracy = \frac{TP + TN}{TP + TN + FP + FN}$$

- **F1 Score:** Precision and Recall are interrelated, Precision and recall are inversely related to each other. Which metrics to use for evaluation depends upon the problem statement. Hence, another solution to evaluate the model is the F1 score, which utilizes both Precision and recalls to obtain the weighted average of Precision & recall. This Metric is used in those cases where class distribution is nonuniform.

$$F1\ Score = 2 \times \frac{Precision \times Recall}{Precision + Recall}.$$

## 3.7 Tools and Techniques Used

To carry out the development of the model we used several tools.
- **ADB:** ADB [21] is a cmd line tool used to communicate with an android device at runtime. It is a client-server architecture with three components a client, a demon, and a server
- **MonkeyRunner:** MonkeyRunner [9] is a testing tool for android devices. We can test programs in python.
- **Jupyter Notebook:** The Jupyter Notebook [22] is a web-based tool used to manage computational Documents. We can do coding, debugging, and analysis in python. It also supports rich media output.
- **Strace:** Strace [23] is a cmd line tool used to trace the system call in the Linux system.
- **Aapt:** Aapt [24] stands for android asset packaging tool. It is a cmd line tool used to view, create and update apk files.
- **Grid Search:** Grid search is a process that searches optimal hyperparameter from a manually specified subset for the targeted algorithm
- **K- Fold Method:** K-fold is a cross-validation technique to evaluate a machine learning model on a very limited dataset. In this technique, data is divided into k parts where k -1 parts are used for training and one for evaluating. The average result of k parts is used as the final results

## 4. Proposed Approach for Detection and Classification of Android Malware

The work done in this research focuses on detecting and classifying android malware using system-level behavior. There are several reasons to use the system-level behavior to detect android malware [25]. Firstly, the system calls are the only way of communication for an application to access the operating system-related services of Android. Therefore, any malicious application cannot deal with sensitive data without system calls. Secondly, the system calls can be tracked without knowing the application. However, system calls represent the low-level behavior of the application [26], and therefore, to overcome this gap, we use machine learning models to match the patterns of different categories of apps.

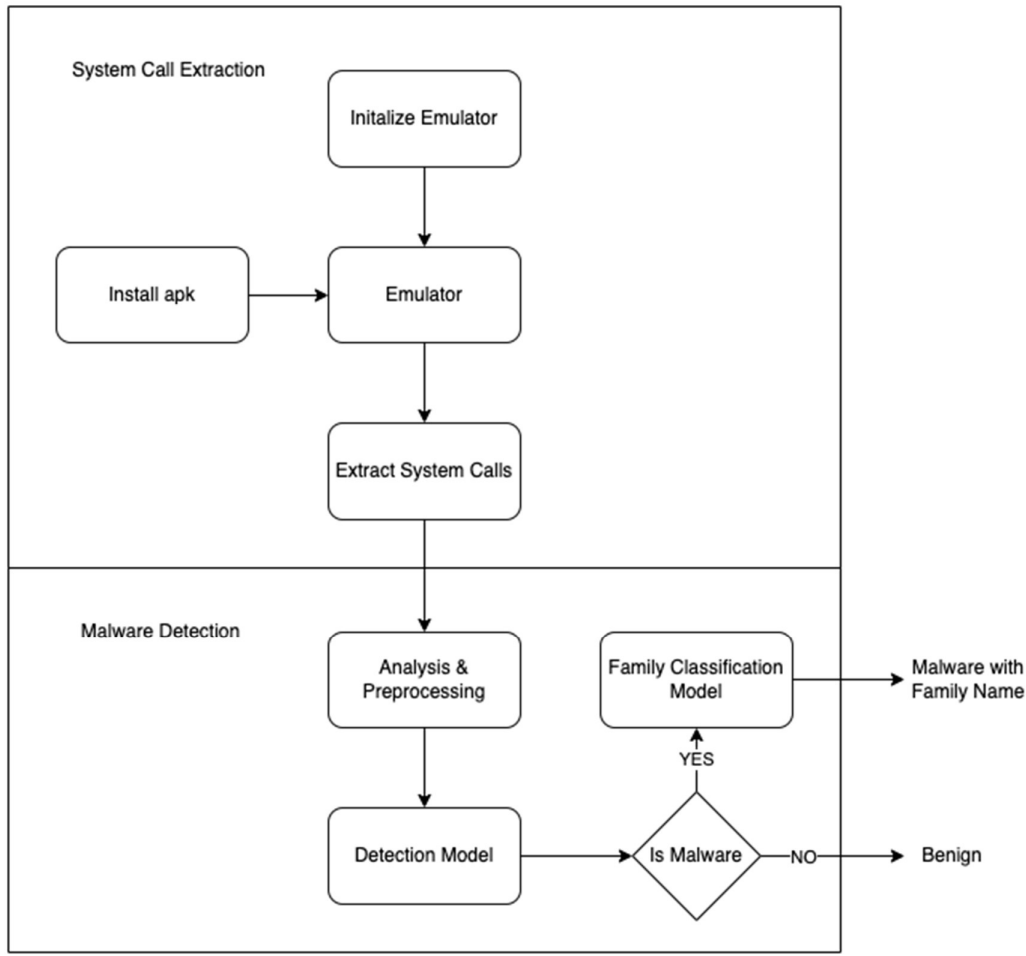

Fig 3.1 System Architecture of Malware Detection
and Family Classification using System Calls

The primary aim of this work is to develop a solution that can effectively perform dynamic analysis of android malware on a large scale. Figure 3.1 shows the architecture of our approach by effectively extracting system calls and making results out of the data.

### 4.1 System Call Extraction

This section contains an approach to collecting execution traces of apk files by constantly interacting with an automation tool, Monkey [9]. We are approximating to minimize the gap between user experience and automation tools for apk. This approach also contains a list of common system events knowns as intent [27] in android to trigger the malware. A shell script that interacts with ADB performs all this functionality with other tools like Monkey, telnet, etc. This script follows a procedure as below:

1. Initialize the saved Snapshot inside the emulator and wait for the system to reach its stable state (check for boot status).
2. Check for emulator connection using monkey
3. Extract the package name from apk (using the AAPT tool)
4. Get root access for the current session

5. Install the apk with all permissions and run the pkg with the monkey tool.
6. Wait for the app to reach its stable state
7. Extract PID using package name and attach Strace with the current PID
8. Send random interaction events to the app using the Monkey tool.
9. Send the selected simulated system event to the app from the list
10. Perform step 8 repeatedly until all system events are considered
11. Kill the app using PID
12. Pull out the log file of Strace and Uninstall the app.

In the above procedure to extract system calls, a list of system events is repeatedly sent to the app. These System events [27] are the standard activity action provided by the android to communicate between android systems. These events can be sent to a process via using the ADB tool. Some vital system events triggering the malicious behavior [13] are BOOT_COMPLETE, ACTION_ANSWER, NEW_OUTGOING_CALL, etc. BOOT event is the most common among to activate the malware. However, the script uses a long list of system events mentioned in Table 3.1.

**Table 3.1:** List of System Events considered for payload Activation

| System Events | |
|---|---|
| BOOT_COMPLETED | DEVICE_STORAGE_OK |
| SMS_RECEIVED | DEVICE_STORAGE_LOW |
| DATA_SMS_RECEIVED | SEND_TO |
| ACTION_ANSWERE | SMS_FULL |
| NEW_OUTGOING | SMS_SERVICE |
| ACTION_POWER_CONNECTED | STATE_CHANGED |
| ACTION_POWER_DISCONNECTED | AIRPLANE_MODE |
| BATTERY_OKAY | BATTERY_CHANGED |
| BATTERY_LOW | DATE_CHANGED |
| BATTERY_EMPTY | INPUT_METHOD_CHANGED |
| CONFIGURATION_CHANGED | PROXY_CHANGE |
| WAP_PUSH_RECEIVED | INPUT_METHOD_CHANGED |

The pulled-out log file of system calls contains data in a table structure, which contains five columns: name of system call (syscall), calls, errors, time in seconds(seconds), percentage of time(%time), and columns with time in sec/total no of calls (usecs/call) as shown in Fig 3.2. This output is from the log file a tabular structure received in the text format.
 It may be observed that some of the values in time are zero, but there is more significance present than having the zero values because these calls consume a total time of less than one microsecond. The last row contains a total of all rows.

```
 1 % time     seconds  usecs/call     calls    errors syscall
 2 ------ ----------- ----------- --------- --------- ----------------
 3  44.66    0.202104          96      2097         1 epoll_pwait
 4  24.58    0.111245          65      1718       347 futex
 5  12.68    0.057404          31      1876           write
 6   8.58    0.038839          11      3535      1869 recvfrom
 7   4.53    0.020520          10      1971           ioctl
 8   3.11    0.014076           7      2099           getuid
 9   0.77    0.003468           3      1139           sendto
10   0.74    0.003331         303        11           madvise
11   0.23    0.001042           1       906           fstat
12   0.06    0.000264           1       494           writev
13   0.03    0.000154           1       152           pread64
14   0.02    0.000105           0       564           read
15   0.00    0.000000           0        76           rt_sigprocmask
16   0.00    0.000000           0        10           sched_yield
17   0.00    0.000000           0        18           dup
18   0.00    0.000000           0         6           clone
19   0.00    0.000000           0        18           fcntl
20   0.00    0.000000           0        18           prctl
21   0.00    0.000000           0        35           epoll_ctl
22   0.00    0.000000           0        11           openat
23   0.00    0.000000           0        47           close
24   0.00    0.000000           0       100           mmap
25   0.00    0.000000           0        12           mprotect
26   0.00    0.000000           0        76           munmap
27 ------ ----------- ----------- --------- --------- ----------------
28 100.00    0.452552                 16989      2217 total
```

Figure 3.2 Extracted strace Log file

## 4.2 Dataset

In our dynamic analysis and detection of malware, we need both benign and malware apk files diverse in malware to classify their behavior. To fulfill this purpose, we used **CIC-ANDMAL2017** [28]. This dataset contains four different categories of malware with ten different malware of each type: **Adware**, **Ransomware**, **Scareware**, and **SMS Malware,** with a total of 426 malware. This dataset also contains 500 benign samples downloaded from Google Play Market. However, these numbers indicate the apk file. In reality, when we pass these apps through our system-call extractor, some apps cannot give desired results due to the compatibility of apk with the current system at various stages. We choose the best settings of the current system mentioned in the next section of this section so that it can support the maximum number of apps. The final count of each sample is discussed in table 3.2

**Table 3.2:** Number of Log files of each category

| Benign | | Malicious | | | |
|---|---|---|---|---|---|
| | | Adware | Ransomware | Scareware | SMS Malware |
| No. of Log Files | 324 | 35 | 78 | 72 | 96 |

## 4.4 Emulation

Google Android Emulator provides a simulation of an actual android device [29]. Our configuration of the emulator is set up to provide any app, either malware or benign very similar environment to a real android phone. Due to compatibility issues of apps with the android version the emulator configuration is set such that it can extract maximum log files from our dataset. Our emulator supports Android 8.0 (Oreo) with 4GB of Ram and high all settings that can be provided because some apps intentionally or unintentionally use high computation; we want all the traces. We also set up a Google account because many of the basic system app functionality is enabled only with a Google account.

## 5. Experiment & Evaluation

In this section of the paper, we describe the experiment we performed to evaluate our approach using Machine Learning Models. Analyze the extracted system calls, gaining some insights from data before detecting malware and changing data into a suitable structure to feed to Machine Learning Models, and evaluate the results of Detection models and classification models.

### 5.1 Analysis & Preprocessing of System Calls

In this preprocessing phase, all collected system calls from the different apps are collectively analyzed with their category. In the tables, some salient information about the system call traces is shown. Table 4.1 represents the average frequency of system calls of the app. From the data, it can be observed that the malware category frequency is lower than the benign apps. This shows that malware's behavior is minimal compared to the benign app. This behavior is pronounced because malware is designed for specific actions.

**Table 4.1:** Average Frequency of Calls and Error

|  | Call | Error | Error % |
|---|---|---|---|
| **Malware** | 19112 | 1852 | 9.69 |
| **Benign** | 35576 | 4601 | 12.93 |

Table 4.2 shows the ten most called system calls in malware and benign sample from this table. We can see some topmost system calls of both categories are the same. Although **ioctl, read** and **write** show some exceptions in the malware category, which indicates malware most frequently calls ioctl, a system call for device control. Also, the frequency of reading and writing is higher in malware apps. Other than this,
We can see **epoll_pwait** is among the topmost system call, which indicates that our approach to system call extraction makes the system wait for input.

**Table 4.2**: Top 10 system calls of each category
based on the average frequency of system calls per android app

| Malware | | Benign | |
|---|---|---|---|
| System call | Percentage | System call | Percentage |
| recvform | 14.37 | Recvform | 19.83 |
| ioctl | 11.7 | write | 11.09 |
| epoll_pwait | 11.46 | futex | 10.49 |
| read | 10.63 | epoll_pwait | 10.38 |
| getuid | 9.82 | ioctl | 8.92 |
| write | 9.2 | getuid | 8.21 |
| futex | 6.36 | sendto | 7.43 |
| sendto | 5.71 | read | 6.33 |
| writev | 2.73 | writev | 3.35 |

We get all raw data direct from the emulator apart from making insight from that data, we need to structure it in a suitable way that can be fit for input for models. We get our data in the form of a row and columns to change it into an input feature. We treat every call and error as one of the features, and the total number of calls made by that particular system call is the value. We extracted 95 different system calls from our log data. This number of extracted system calls can be different for other datasets because we considering only those system calls, which are invoked by any of the applications in the dataset. Considering other unused system calls will only increase the input feature vector and size of the model which will result in more computation

In total, we have 190 features, half from the system calls and half from errors of system calls. Although we have other columns in log files, they are not distinctive as considered one. To reduce computation costs at later stages, we normalized our data on a zero to one scale by using the total number of calls and errors.

## 5.2 Malware Detection & Family Classification

In this section, we discuss the model used for classification. This paper used five different models for both malware detection and family classification. In malware detection, KNN gave the best result with an F1 score of 0.85 on unseen data, and the Decision tree top with an F1 score equal to 0.73 for family classification. We used the grid search technique on the training data set to optimize the parameters for better results with the k-fold method as cross-validation. We used an average of all results concerning that parameter to analyze the relative performance of the model. The whole data set is divided into two parts with 1/5th sample for testing and the remaining for training.

**Table 4.3:** Average accuracy of models with respect to parameters

| | Parameter | | Average Accuracy | |
|---|---|---|---|---|
| Model | Name | Value | Detection | Family Classification |

| Model | Hyperparameter | Value | | |
|---|---|---|---|---|
| **Decision Tree** | criterion | entropy | 76.01 | 59.71 |
| | | gini | 77.08 | 58.54 |
| | Max Depth | 2 | 72.27 | 41.29 |
| | | 5 | 76.15 | 54.69 |
| | | 8 | 78.59 | 61.16 |
| | | 10 | 77.62 | 63.84 |
| | | 30 | 77.32 | 63.17 |
| | | 50 | 76.84 | 64.29 |
| | | 100 | 77.03 | 62.5 |
| | | 200 | 76.55 | 62.05 |
| **KNN** | N neighbors | 3 | 79.69 | 66.22 |
| | | 5 | 80.27 | 62.5 |
| | | 7 | 79.55 | 59.08 |
| | | 11 | 78.71 | 48.96 |
| | Distance Type | Manhattan | 80.18 | 59.93 |
| | | Euclidean | 79.36 | 59.49 |
| | | Minkowski | 79.12 | 58.15 |
| **Logistic Regression** | C | 0.01 | 61.57 | 40.18 |
| | | 0.1 | 65.67 | 50.45 |
| | | 1 | 73.45 | 57.14 |
| | | 10 | 74.23 | 60.27 |
| | | 100 | 75.69 | 64.55 |
| | | 1000 | 76.2 | 65.74 |
| | Solver | lbfgs | 64.17 | 54.64 |
| | | liblinear | 65.32 | 56.77 |
| | | newton-cg | 72.52 | 56.85 |
| | | sag | 70.45 | 56.1 |
| | | saga | 68.41 | 55.73 |
| **Support Vector Classifier** | C | 0.1 | 58.56 | 52.01 |
| | | 1 | 66.76 | 52.31 |
| | | 10 | 68.54 | 54.69 |
| | | 100 | 71.85 | 59.23 |
| | | 1000 | 72.66 | 62.2 |
| | | 10000 | 74.73 | 65.55 |
| | Kernel | linear | 72.3 | 60.19 |
| | | poly | 64.77 | 54.91 |
| | | rbf | 69.48 | 57.89 |
| | Gamma | auto | 64.27 | 54.44 |
| | | scale | 73.43 | 60.89 |
| **MLP** | alpha | 0.0001 | 76.17 | 60.11 |
| | | 0.005 | 76.4 | 60.42 |
| | | 0.05 | 75.39 | 59.78 |
| | | 0.1 | 74.75 | 59.24 |
| | | (100, 100) | 66.07 | - |
| | | (100, 100, 100) | 66.19 | - |
| | | (100, 200) | 65.88 | - |
| | | (100, 200, 100) | 65.97 | - |
| | Hidden Layers | (200, 100) | - | 59.38 |
| | | (200, 200) | - | 61.64 |
| | | (200, 200, 100) | - | 66.49 |
| | | (200, 200, 200) | - | 66.35 |
| | | (200,-) | - | 45.59 |
| | Iterations | 200 | 65.61 | 51.21 |
| | | 400 | 65.98 | 59.33 |
| | | 600 | 66.6 | 63.42 |

Parameters from each model with the best results are used to evaluate test samples. For example, in Support Vector Classifier Linear Kernel with scaled gamma and C equal to 10000 is used to evaluate the Test set.

Different methods of evaluating a model indicate or convey different information. To evaluate family classification, precision is the optimal choice because it will tell us how many times it is predicting that particular family correctly. But in the case of malware detection recall should be considered because it conveys how many times we identified malware correctly. Apart from precision and recall, the F1 score is the metric that evaluates the whole model as complete because it utilizes both precision and recall. In our case of malware detection best recall value of 0.85 is given by K Nearest Neighbour Model and for family classification, the best precision value of 0.74 is given by Decision Tree Model.

**Table 4.4:** Precision, Recall, and F1 Score of Detection and Family Classification on test samples

| | Precision | | Recall | | F1 Score | |
|---|---|---|---|---|---|---|
| | Detection | Family Classification | Detection | Family Classification | Detection | Family Classification |
| **Logistic Regression** | 0.84 | 0.72 | 0.84 | 0.65 | 0.84 | 0.68 |
| **KNN** | 0.85 | 0.67 | 0.85 | 0.61 | 0.85 | 0.63 |
| **Support Vector Classifier** | 0.8 | 0.71 | 0.79 | 0.68 | 0.79 | 0.7 |
| **Decision Tree** | 0.72 | 0.74 | 0.73 | 0.74 | 0.72 | 0.73 |
| **MLP** | 0.74 | 0.71 | 0.73 | 0.68 | 0.73 | 0.69 |

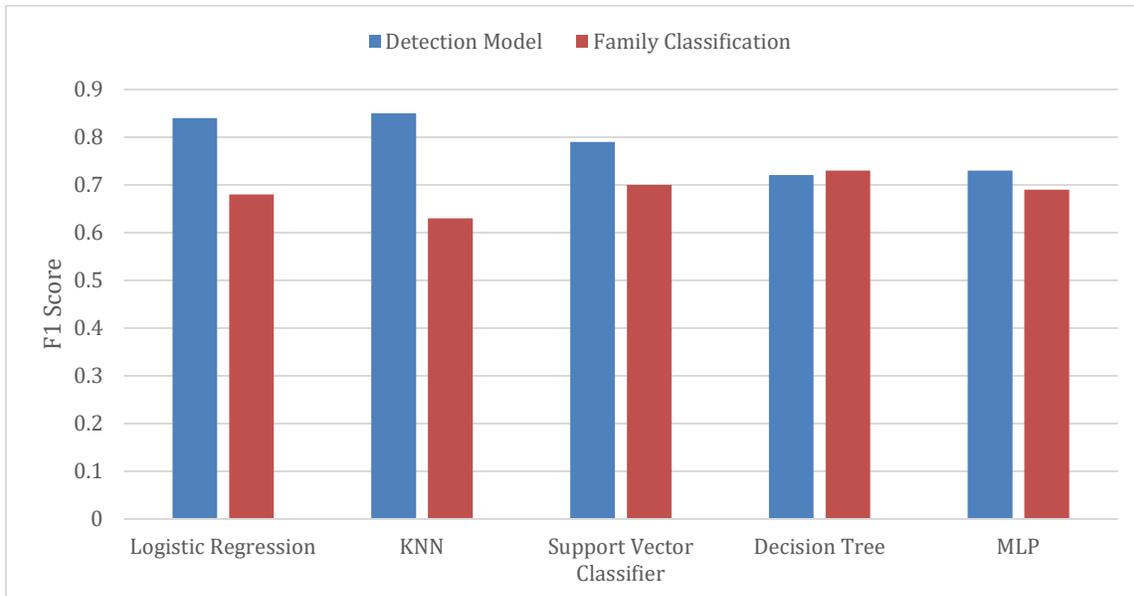

**Fig 4.1:** F1 Score of Malware Detector Models and Family Classifier Model

## 6. Conclusion

With swift, dynamically changing android, android malware is adopting this change too. We present a method to implement this automatic extraction of the runtime behavior of the android app with reduced the gap between the human experience of malware and automation tools in a monitored environment. This technique can be scaled to complete automation of extracting the runtime behavior of apps. Moreover, tracing app behavior at a low level is hard to beat without detecting its signature at a low level. We used machine learning to match the system-level app behavior with low-level extracted data. With a small set of apps, we got an F1 Score up to 0.85 in detection and 0.73 in family classification. In future work, we would like to investigate the following points:
- Making automation techniques to extract runtime behavior more simulated to human experience
- Scaling the current technique on a large dataset with more variety of malware
- Consider other runtime features like network package and API tracing, etc. for the input feature